\documentclass[12pt]{article}
\usepackage{a4wide}
\usepackage{axodraw}
\usepackage{graphicx}

\newcommand{\be}{\begin{equation}}
\newcommand{\ee}{\end{equation}}
\newcommand{\ba}{\begin{eqnarray}}
\newcommand{\ea}{\end{eqnarray}}
\newcommand{\eq}{\begin{equation}}
\newcommand{\en}{\\[2mm]\end{equation}}
\newcommand{\bea}{\begin{eqnarray}}
\newcommand{\ena}{\end{eqnarray}}
\newcommand{\order}[1]{\ensuremath{{\cal O}(#1)}}
\begin{document}
\begin {titlepage}
\begin{flushright}
\rm LU TP 01-01\\
IPNO-DR/01-003\\
hep-ph/0101127\\
January 2001
\end{flushright}
\vfill
\begin{center}
{\Large\bf
QCD Isospin Breaking in Meson Masses, Decay Constants and Quark Mass Ratios
}\\[1cm]
\vfill
{\bf G.~Amoros$^{a,b}$, J. Bijnens$^a$ and P. Talavera$^c$}\\[0.5cm]

{$^a$ Department of Theoretical Physics 2, Lund University,\\
S\"olvegatan 14A, S22362 Lund, Sweden\\[0.3cm]

$^b$ Departament de F{\'\i}sica Te\`orica, IFIC, Universitat de
Val\`encia-CSIC,\\
Apt. Correus 22085, E-46071 Val\`encia, Spain\\[0.3cm]

$^c$Groupe de Physique Th\'eorique,
Institut de Physique Nucl\'eaire,\\
Universit\'e Paris Sud,
F-91406 Orsay Cedex, France}
\vfill

\begin{abstract}
The procedure to calculate masses and matrix-elements in the presence
of mixing of the basis states is explained in detail. We then apply
this procedure to the two-loop calculation in Chiral Perturbation Theory
of pseudoscalar masses and decay constants including quark mass
isospin breaking. These results are used to update our analysis
of $K_{\ell4}$ done previously and obtain a value of
$m_u/m_d$ in addition to values for the low-energy-constants $L_i^r$.
\end{abstract}

\vfill
\end{center}
\vfill
\end{titlepage}

\section{Introduction}

Experiments in low-energy QCD are becoming more and more precise.
It is therefore needed that theoretical calculations
are performed to similar accuracy. In the
purely mesonic sector the low-energy effective theory of QCD is known
as an effective theory in terms of pseudo-scalar mesons only.
It is known as  Chiral Perturbation Theory (CHPT)
and was put on a solid theoretical basis in \cite{GL}.
In the two-flavour sector calculations to two-loop order are now
customary. In our earlier work \cite{ABT1,ABT2}, where also references to other
calculations can be found, we have performed the main calculations
in the three flavour sector also to two loops in the isospin symmetric
approximation.

One of the remaining uncertainties was the importance of isospin breaking
effects. Basically in our previous work and many other
theoretical calculations
isospin breaking
was only included by guessing its effect and putting it in the
uncertainty of the final result.
An example where we definitely need to
go beyond this is in the use of
$K_{\ell3}$ form-factors \cite{roos} to extract an accurate value
of the $\vert\, V_{us}\,\vert$ CKM matrix-element.
Here we take a first step in that we 
evaluate within CHPT
the strong isospin breaking in masses and decay constants to
next-to-next-to-leading order. 

This work serves two purposes: it checks the
dependence on isospin breaking of the determination of the CHPT 
low-energy constants (LEC's) at \order{p^4} performed in our
earlier work \cite{ABT1,ABT2} and allows us to extract information
on the quark mass ratio $m_u/m_d$. The latter is one of the fundamental
input parameters in QCD and thus needs to be determined as accurately
as possible.

Another motivation is to see how our previous results \cite{ABT1,ABT2}
can change with the new experimental data on $K_{\ell4}$
\cite{truol}. In particular, they contain one more independent
measurement, the slope of the $G$ form-factor, allowing for a cross-check
on the CHPT description and we will use it to have a first check
on the large $N_c$ assumptions made in the theoretical analysis.

The plan of the paper is the following\,:
We describe in Sect.~\ref{formalism} the changes needed
in the presence of mixing of the external states.
We first discuss it for the masses and second for matrix-elements.
We then sketch all the computed quantities,
Sect.~\ref{quantities}, and follow it with a discussion
of the electromagnetic corrections in Sect. \ref{em}.
The latter remain a sizable source of uncertainty in our results.
In Sect.~\ref{assuptions} we list all the assumptions on which the calculation
relies. Sect.~\ref{data} introduces one of the main inputs, 
the $K_{\ell 4}$ form-factors, discussing
the main differences in the data and parametrization 
used in this analysis w.r.t. the previous one.
Sect.~\ref{fitting}
 presents a brief summary of our fitting procedure with
all the inputs and output variables. Sect.~\ref{oldres}
is devoted to the update of 
the results presented already in \cite{ABT1,ABT2}. In Sect.~\ref{ratios}
we discuss
the impact of the outputs on the quark mass ratios. And finally we
briefly summarize our findings in Sect.~\ref{sum}.

The formulas are of such a length that they cannot be presented in
a manuscript of reasonable length and we have therefore not included them.
For an introduction to CHPT we refer to \cite{CHPTlectures}.

\section{Formalism}
\label{formalism}

In the presence of mixing, where each external field can couple to more
than one-particle state, we have to generalize the discussion
of masses and amplitudes given in \cite{ABT1} somewhat.

We first discuss how the masses can be obtained from a general
two-point function and later the
generalization of the wave-function-renormalization part for matrix-element
calculations. The notation will be chosen appropriate
for the $\pi^0$-$\eta$ mixing case, where the inverse of a two-by-two
matrix can be written explicitly in a simple manner.

\subsection{Masses}

\begin{figure}
\label{1pi}
\begin{center}
\setlength{\unitlength}{1.65pt}
\begin{picture}(240,30)(-30,23)
\SetScale{1.65}               
\SetWidth{1.0}
\Line(-30,40)(-10,40)
\Text(-28,43)[b]{i}
\Text(-12,43)[b]{j}
\Text(-3,40)[]{+}
\Line(4,40)(20,40)
\Text(5,43)[b]{i}
\CCirc(25,40){5}{Black}{Yellow}
\Line(30,40)(46,40)
\Text(44,43)[b]{j}
\Text(50,40)[]{+}
\Line(54,40)(70,40)
\Text(56,43)[b]{i}
\CCirc(75,40){5}{Black}{Yellow}
\Line(80,40)(90,40)
\CCirc(95,40){5}{Black}{Yellow}
\Line(100,40)(116,40)
\Text(114,43)[b]{j}
\Text(120,40)[]{+}
%
\Line(124,40)(140,40)
\Text(126,43)[b]{i}
\CCirc(145,40){5}{Black}{Yellow}
\Line(150,40)(160,40)
\CCirc(165,40){5}{Black}{Yellow}
\Line(170,40)(180,40)
\CCirc(185,40){5}{Black}{Yellow}
\Line(190,40)(206,40)
\Text(204,43)[b]{j}
\end{picture}
\\
\setlength{\unitlength}{1.65pt}
\begin{picture}(145,20)(-35,15)
\SetScale{1.65}               
\SetWidth{1.0}
\Text(-33,20)[]{+}
\Vertex(-27,20){1}
\Vertex(-23,20){1}
\Vertex(-19,20){1}
\Vertex(-15,20){1}
\Text(-5,20)[]{+}
\Line(0,20)(10,20)
\Text(2,23)[b]{i}
\CCirc(15,20){5}{Black}{Yellow}
\Line(20,20)(30,20)
\CCirc(35,20){5}{Black}{Yellow}
\Line(40,20)(45,20)
\Vertex(49,20){1}
\Vertex(53,20){1}
\Vertex(57,20){1}
\Vertex(61,20){1}
\Line(65,20)(70,20)
\CCirc(75,20){5}{Black}{Yellow}
\Line(80,20)(90,20)
\CCirc(95,20){5}{Black}{Yellow}
\Line(100,20)(110,20)
\Text(108,23)[b]{j}
\end{picture}
\end{center}
\caption{\label{figTWOP}
The diagrams contributing to the
two-point function $G_{ij}$.
The filled circles correspond to one-particle-irreducible
diagrams and the lines to lowest order propagators.}
\end{figure}
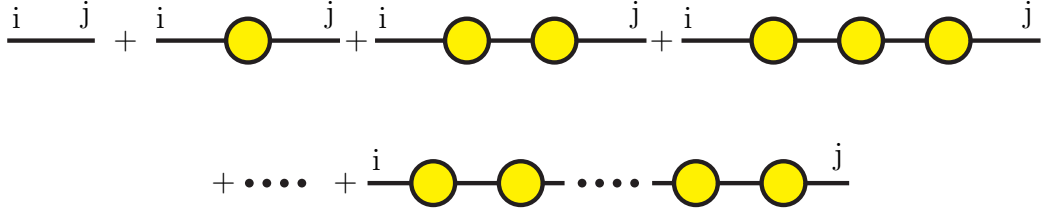

In terms of the lowest order fields $\phi_i$ we define the two-point
function
\be
\label{propagator}
G_{ij}(p^2,m_{i0}^2,F_0) =
\int d^dx e^{i\,p\cdot x}
\langle \,0\, \vert\,T\left\{ \phi_i(x)\, \phi_j(0)\right\}
\, \vert\, 0\, \rangle\,.
\ee
We have suppressed the dependence on everything except $p^2$, $F_0$
and the lowest order masses $m_{i0}$.
The diagrams contributing to this are depicted in Fig.~\ref{figTWOP}.
Writing the lowest order propagator as $i P_{ij}$
and the one-particle-irreducible contributions as $i\Pi_{kl}$,
pictorially $i\Pi_{kl} = $\setlength{\unitlength}{1pt}
\begin{picture}(25,10)(-5,0)\Line(-1,5)(2,5)\CCirc(7,5){5}{Black}{Yellow}
\Line(12,5)(15,5)\Text(-3,5)[]{\scriptsize k}\Text(17,5)[]{\scriptsize l}
\end{picture},
we can obtain the full two-point function in matrix notation as
\be
G = iP + iP(i\Pi iP) + iP(i\Pi iP)^2 + \ldots
= i P \left(1+\Pi P\right)^{-1}\,.
\ee
In the case of no mixing, $\Pi$, $P$ and $G$ are all diagonal matrices
and the solution is simple. The masses of the particles correspond to
the poles which in the diagonal case are given by the zeros of
$p^2-m_0^2-\Pi$ as described in \cite{ABT1}.
In the general case they are again given by the poles in $G$ as a matrix.
These can be most easily found as places where the inverse of $G$
has a zero eigenvalue or the determinant vanishes
\be
\label{determinant}
\det(G^{-1}) = \det\left(\left(1+\Pi P\right)P^{-1}\right)
= \det\left(P^{-1}+\Pi\right) = 0\,.
\ee
It is this equation we now like to solve perturbatively in our case
of $\pi^0$-$\eta$ mixing. The isospin eigenstates used in the Lagrangian
are denoted by $\pi^0_3$ and $\eta_8$. To lowest order, we can diagonalize
$G$ and thus by definition $P$ by a simple rotation
\ba
\label{defpi0eta}
\pi^0 &\equiv& \cos(\epsilon)\, \pi^0_3 + \sin(\epsilon)\, \eta_8\,,\\
\nonumber 
\eta &\equiv& -\sin(\epsilon)\, \pi^0_3 + \cos(\epsilon)\, \eta_8\,.
\ea
where the lowest order mixing angle $\epsilon$ satisfies
\be
\label{angle}
\tan(2\epsilon)
 = \frac{\sqrt{3}}{2}\frac{m_d-m_u}{m_s-\hat{m}}\,,
\ee
with $\hat{m}=(m_u+m_d)/2$, the average of the lightest quark masses.
In the basis defined by $\pi^0$, $\eta$, the Kaon and charged pion
fields, we have
\be
P_{ij} = \frac{\delta_{ij}}{p^2-m^2_{i0}}
\ee
and $\Pi$ starts by definition only at next-to-leading order,
\order{p^4}, but does not need
to be diagonal. The lowest order mixing angle $\epsilon$ does not need to
be small, we will keep it to all orders throughout our calculations.

With the previous prescriptions
$P$ is diagonal and $\Pi$ is blockdiagonal enforced by the various
other conserved quantities like, for instance, parity. 
Then each block can be treated separately.
The charged pion and Kaon do obey a simple diagonal equation without
mixing, therefore only the $\pi^0$-$\eta$ block needs special consideration.

In order to obtain the chiral expansion of the masses, we need
to find the zeros of the determinant in Eq. (\ref{determinant}) as a function
of
\be
\label{defmass}
p^2 = m_{phys}^2 = m_0^2 + (m^2)^{(4)} + (m^2)^{(6)} + \ldots\,,
\ee
where the superindices refer to the chiral order.
In the same way we expand
the one-particle-irreducible 
self interaction, $\Pi(m_{phys}^2,m_{i0}^2,F_0)$,
in chiral orders
\be
\label{pi}
\Pi(m_{phys}^2,m_{i0}^2,F_0) = 
 \Pi^{(4)}(m_{phys}^2,m_{i0}^2,F_0)
 + \Pi^{(6)}(m_{phys}^2,m_{i0}^2,F_0) + \ldots\,.
\ee
Using $3$ and $8$ as labels for the $\pi^0$ and $\eta$ fields defined
in Eq. (\ref{defpi0eta}) and temporarily suppressing the functional dependence
in $\Pi$ the equation becomes to \order{p^6}
\be
\left(m_{phys}^2 - m_{30}^2 + \Pi_{33}^{(4)} + \Pi_{33}^{(6)}\right)
\left(m_{phys}^2 - m_{80}^2 + \Pi_{88}^{(4)} + \Pi_{88}^{(6)}\right)
-
\left( \Pi_{38}^{(4)} + \Pi_{38}^{(6)}\right)^2 = 0\,,
\ee
where the matrix $\Pi$ is symmetric.
We insert at this point the expansion of $m_{phys}^2$ and then either
the lowest order mass is $m^2_{30}$ or $m^2_{80}$. For definiteness hereafter
only the pion mass case will be considered. Thus we obtain
\ba
m_0^2 &=& m^2_{30}\,, \nonumber\\
(m^2)^{(4)} &=& -\Pi_{33}^{(4)}(m_{30}^2, m_{i0}^2, F_0)\,, \nonumber\\
(m^2)^{(6)} &=& -\Pi_{33}^{(4)}(m_{30}^2, m_{i0}^2, F_0)
  - (m^2)^{(4)}\left.\frac{\partial}{\partial p^2}
    \Pi_{33}^{(4)}(p^2, m_{i0}^2, F_0)\right|_{p^2 = m_{30}^2}\nonumber\\
&&  +\frac{1}{m_{30}^2-m_{80}^2}\left( \Pi_{38}^{(4)}(m_{30}^2, m_{i0}^2, F_0)
   \right)^2\,.
\ea
Notice that we have chosen here to use the strict expansion where we always
express the arguments of the selfenergy in lowest order quantities.
For numerical results later we will express all masses in terms of their
physical masses and $F_0$ in terms of $F_{\pi^\pm}$ and reorder
 the series accordingly.

We have only used the fact here that a lowest order mixing angle is
well-defined. There is of course no reason for this to be true at higher
orders and indeed it is not the case already at order $p^4$ \cite{GL}.
When photonic corrections are included, the possibility of
mixing of states with additional soft photons of course introduces
the usual infrared problems and provides another reason for not
having a simple mixing angle for the physical states \cite{helmutmixing}.

Another alternative is to make a second, $p^2$-dependent, field redefinition
such that
$P_{ij}^{-1}+\Pi_{ij}^{(4)}(p^2,m_{i0}^2,F_0)$, $(i\ne j)$ vanishes.
In principle in terms of those fields the expressions 
for the matrix-elements derived below are somewhat simpler.
The additional complications in calculating the diagrams is the reason we
work with fields that are only diagonalized to lowest order.

\subsection{Matrix-elements}

To calculate any observable we follow the usual LSZ reduction formalism.
The matrix-element in momentum space for any
n incoming or outgoing states is
\be
{\cal A}_{i_1\ldots i_n}
= 
\left( \frac{(-i)^{n}}{\sqrt{Z_{i_1} \ldots Z_{i_n}}} \right)\, 
\prod_{i=1}^n \, 
\lim_{k_i^2\to m_i^2}(k_i^2-m_i^2)\,
G_{i_1\ldots i_{n}}
(k_1,\ldots,k_n)\,.
\ee
The function $G_{i_1\ldots i_{n}}
(k_1,\ldots,k_n)$ is the full n-point Green function generated by the
$n$ fields $\phi_{i_1}(k_1),\ldots,\phi_{i_n}(k_n)$.
The coefficients $Z_i$ are defined via
\be
\label{defZ}
G_{ii}(p^2\approx m_{i\,phys}^2)
= \frac{Z_i}{p^2-m_{i\,phys}^2}+\order{1}
\ee
and are often referred to as wave-function-renormalization.
In the diagonal case this is the procedure described
in detail in \cite{pipi}. In the case where mixing is present, the formulas
become more cumbersome.
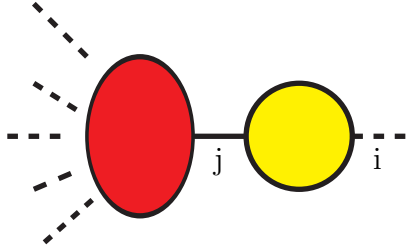
\begin{figure}
\begin{center}
\setlength{\unitlength}{2pt}
\SetScale{2.0}
\SetWidth{1.0}
\begin{picture}(40,30)(10,10)
\COval(25,25)(15,10)(0){Black}{Red}
\DashLine(0,25)(10,25){2}
\DashLine(5,50)(15,40){2}
\DashLine(5,35)(13,30){2}
\DashLine(12,18)(5,15){2}
\DashLine(7,5)(16,13){2}
\Line(35,25)(45,25)
\Text(40,23)[t]{j}
\CCirc(55,25){10}{Black}{Yellow}
\DashLine(65,25)(75,25){2}
\Text(70,23)[t]{i}
\end{picture}
\end{center}
\caption{\label{figMATRIX} A general contribution to
$G_{i_1 i_2 \ldots i_n}$. We choose one of the legs, $i_l = i$, and show
how the full two-point function $G_{ij}$ contributes on that leg,
denoted by the filled circle on the right-hand-side.
The remainder of the diagrams, the Green-function, reduced on the $i_h$
leg is depicted by the oval on the left-hand-side and denoted in the text by
${\cal G}_{i_1 i_2 \ldots i_n}$ with $i_h = j$.}
\end{figure}
We will for simplicity discuss only the case of
one external leg but the generalization
should be obvious. In terms of the Green-function ${\cal G}_j$ which
is one-particle-irreducible and amputated on the $j$-leg which can connect to
the $\phi_i$ external state we then get
\be
{\cal A}_i = \frac{-i}{\sqrt{Z_i}}\lim_{k_i^2\to m_i^2}(k_i^2-m_i^2)
G_{ij}(k_i^2){\cal G}_j
\ee
with an implicit summation over $j$.

The quantity, simply calculable in terms of $\Pi$, was $G^{-1} = -i{\cal P}$,
${\cal P} = P^{-1}+\Pi$.
Now for a two-by-two matrix the inverse is fairly easy
\be
G = G(p^2) = \frac{i}{{\cal P}_{33}(p^2)\,{\cal P}_{88}(p^2)
 - {\cal P}_{38}^2(p^2)}
\left( \begin{array}{cc}
{\cal P}_{88}(p^2)   & -{\cal P}_{38}(p^2)  \\
-{\cal P}_{38}(p^2)  & {\cal P}_{33}(p^2) 
    \end{array} \right)\,,
\ee
where we have indicated the $p^2$ dependence explicitly.
At the physical mass $\det{\cal P}(p^2)$ vanishes and thus 
the pole in Eq. (\ref{defZ}) appears naturally. For the $\pi$ and $\eta$ case
we obtain
\be
Z_3 = \frac{1}
{\left.\frac{\partial}{\partial p^2}
\left(\det{\cal P}(p^2)\right)\right|_{p^2=m_\pi^2}} {\cal P}_{88}(m_\pi^2)\,,
\quad
Z_8 = \frac{1}
{\left.\frac{\partial}{\partial p^2}
\left(\det{\cal P}(p^2)\right)\right|_{p^2=m_\eta^2}} 
{\cal P}_{33}(m_\eta^2)\,.
\ee
The final normalized amplitude for the pion case is then
\be
\label{amplitude}
{\cal A}_3 = \frac{1}{\sqrt{ {\cal P}_{88} \frac{\partial}{\partial p^2}
\left( {\cal P}_{33}\,{\cal P}_{88} - {\cal P}_{38}^2 \right) }}
\, \left\{ {\cal P}_{88} {\cal G}_3
- {\cal P}_{38} {\cal G}_8 \right\}\,.
\ee
all functions are taken at the pion mass.
${\cal G}_j$ refers to the l.h.s. of Fig. \ref{figMATRIX},
the dark (red) bubble, which contains
all the one-particle-irreducible in leg $j$ diagrams.

The amplitude in Eq.~(\ref{amplitude}) contains formally all the chiral orders.
Obviously for practical purpose one has to restrict to a given order in the 
chiral series. Therefore we expand ${\cal P}$ and
 $G_i$ inside Eq.~(\ref{amplitude}) up to 
\order{p^6},
but we do not expand these functions in $\epsilon$.

In our basis we have that $(P^{-1})_{ij} = \delta_{ij}(p^2-m_{i0}^2)$
and $\Pi$ only starts at \order{p^4}
which allows to evaluate the expansions explicitly in terms of
\ba
{\cal G}_j &=& {\cal G}_j^{(2)}+{\cal G}_j^{(4)}+{\cal G}_j^{(6)}\,,
\nonumber\\
\Pi_{ij} &=& \Pi_{ij}^{(4)}+\Pi_{ij}^{(6)}\,,
\nonumber\\
Z_{ij} &\equiv&\frac{\partial}{\partial p^2}\Pi_{ij}
= Z_{ij}^{(2)}+Z_{ij}^{(4)}+Z_{ij}^{(6)}\,.
\ea
Using the fact that $\det {\cal P}=0$ to simplify further the expression, 
one explicitly obtains for the pion case
\ba
{\cal A}_3 &=& {\cal G}_3^{(2)} + \bigg\{ 
{\cal G}_3^{(4)} -\frac{1}{2} Z_{33}^{(4)} {\cal G}_3^{(2)}
 -\frac{\Pi_{38}^{(4)}}{\Delta m^2} {\cal G}_8^{(2)}
 \bigg\} 
+\biggl[
{\cal G}_3^{(6)}
-\frac{1}{2} Z_{33}^{(6)} {\cal G}_3^{(2)}
\nonumber\\&&
-\frac{1}{2} Z_{33}^{(4)} {\cal G}_3^{(4)}
+\frac{3}{8}\left( Z_{33}^{(4)}\right)^2 {\cal G}_3^{(2)}
+\frac{Z_{38}^{(4)}\Pi_{38}^{(4)}}{\Delta m^2} {\cal G}_3^{(2)}
-\frac{1}{2}\left(\frac{\Pi_{38}^{(4)}}{\Delta m^2}\right)^2 {\cal G}_3^{(2)}
\nonumber\\&&
-\frac{\Pi_{38}^{(4)}}{\Delta m^2}{\cal G}_8^{(4)}
-\frac{\Pi_{38}^{(6)}}{\Delta m^2}{\cal G}_8^{(2)}
+\frac{\Pi_{38}^{(4)}\,\Pi_{88}^{(4)}}{\Delta m^2}{\cal G}_8^{(2)}
+\frac{1}{2}Z_{33}^{(4)}\frac{\Pi_{38}^{(4)}}{\Delta m^2}{\cal G}_8^{(2)}
\biggr]\,,
\ea
where the terms displayed between curly brackets are of \order{p^4} and
the ones between squared brackets are \order{p^6}. 
The formulas are at $k_i^2=m_\pi^2$, the physical pion mass and
$\Delta m^2 \equiv \left(P^{-1}\right)_{88}(m_{\pi\,phys}^2)=
 m_{\pi\,phys}^2-m_{\eta0}^2$.
We can
then rewrite the dependence on the lowest order masses and decay constants
as before as well in the physical masses and decay constants.

We checked our results in an analytical and numerical way\,: in terms of the
general functions $\sin(\epsilon)$ and $\cos(\epsilon)$ we find a
scale-independent result, while setting $\epsilon=0$ we recover the results
of \cite{ABT1}.

\section{Analytical results}
\label{quantities}

In \cite{ABT1} we presented the masses and decay constants and
in \cite{ABT2} in addition the vacuum expectation values and the
$K_{\ell4}$ form-factors at next-to-next-to-leading order. 
Even in the latter case we had to resort to numerical
approximations for a large part of the expressions involving integrals
in order to keep
formulas to a reasonable size. In the present case, due to the presence of
isospin breaking effects even the expressions for the masses 
become very cumbersome. In this section we therefore only define the quantities
we have calculated and how the various chiral orders have been split.

We have followed as much as possible the philosophy of rewriting the
quark masses in terms of physical meson masses at \order{p^4},
with the exception
of $m_u-m_d$ and the lowest order mixing angle $\epsilon$. We have
however used Eq. (\ref{angle}) as much as possible to reduce the number
of terms. In addition the lowest order mass relations are
extensively used at \order{p^6}.

The quantities calculated up to \order{p^6} including isospin breaking
due to the quark masses to all orders are
\begin{itemize}
\item The 5 meson masses, $m_{\pi^\pm}^2$, $m_{\pi^0}^2$, $m_{\eta}^2$,
$m_{K^\pm}^2$ and $m_{K^0}^2$.
\item The decay constants $F_{\pi^\pm}$, $F_{K^\pm}$ and $F_{K^0}$.
\item The 4-decay constants in the neutral sector.
These are the couplings of the physical $\pi^0$ and $\eta$ states to
the triplet and octet axial current, labeled as $F_{\pi^0 3}$,
$F_{\pi^0 8}$, $F_{\eta 3}$ and $F_{\eta 8}$.
In addition we define the quantities $\epsilon_\pi = F_{\pi^0 8}/F_{\pi^0 3}$
and $\epsilon_\eta = - F_{\eta 3}/F_{\eta 8}$.
\item The three vacuum expectation values $\langle\bar u u\rangle$,
$\langle \bar d d\rangle$ and $\langle \bar s s\rangle$.
\end{itemize}

\section{Treatment of electromagnetic corrections}
\label{em}

Isospin breaking has two different sources, the quark mass difference $m_u-m_d$
and the electromagnetic interaction. The former breaking introduces
only one new parameter as compared to the isospin symmetric analysis of
\cite{ABT1}, $m_u/m_d$. The latter can in principle also be treated within
the framework of CHPT. The corrections to the masses were treated in
\cite{Urech} and the framework needed for the decay constants
and $K_{\ell4}$ was worked out in \cite{KNRT}. Unfortunately, the general
treatment brings in a large number of new unknown constants. Estimating
these brings into the game the evaluation of nonleptonic matrix-elements,
a problem that is beyond our present scope.

The lowest order is \order{e^2 p^0} and brings in only one more parameter
which can be fixed by the difference $m_{\pi^\pm}^2-m_{\pi^0}^2$. 
The quark mass contribution
to this mass difference is very small and can be neglected as a good
first approximation. This leads to the prediction
\be
\label{dashen}
\left(m_{K^\pm}^2-m_{K^0}^2\right)_{\mbox{\small{em}}}
= m_{\pi^\pm}^2-m_{\pi^0}^2\,,
\ee
a result often referred to as Dashen's theorem \cite{Dashen}.
At this order there are no corrections to the decay constants.

At the next chiral order, \order{e^2p^2}, 
there are corrections to Eq. (\ref{dashen}).
The chiral logarithms are of the expected size \cite{Urech,NR}. 
The earlier indications of a large contribution also
from the \order{e^2p^2} LEC's can be found in \cite{hans,dono}. 
These results, but
in a less pronounced manner,
were confirmed by the lattice calculations of \cite{lat}. In what follows
we will 
use the results of \cite{BP} which are compatible
with the averaged value of \cite{hans,dono} 
since it incorporates both a correct
long-short-distance-matching\footnote{It is precisely this matching that
invalidates the results of \cite{BaurUrech}.}
and the chiral logarithms at leading
and nonleading order in $1/N_c$. It also agrees with the results of
 \cite{lat}. The final result of \cite{BP}
was
\be
\label{BPdashen}
\left(m_{K^\pm}^2-m_{K^0}^2\right)_{\mbox{\small{em}}}
= (1.84\pm0.25)\left(m_{\pi^\pm}^2-m_{\pi^0}^2\right)\,.
\ee
We will use the central value of Eq. (\ref{BPdashen}) in our numerical
results and Eq. (\ref{dashen}) to show the dependence of the result
on this quantity. Numerically the above two correspond to
\be
\left(m_{K^\pm}-m_{K^0}\right)_{\mbox{\small{em}}}
\approx 2.3~\mbox{MeV}~\mbox{[Eq. (\ref{BPdashen})]}\quad\mbox{and}\quad
1.3~\mbox{MeV}~
\mbox{[Eq. (\ref{dashen})]}\,.
\ee
We calculate the masses and decay constants with the strong meson masses,
i.e. with the electromagnetic effects above removed.

The electromagnetic corrections to the decay constants are only partially
treated within CHPT. We will use the values as calculated
by \cite{Holstein,Finkemeier} as $F_{\pi^\pm} = 92.4$~MeV
and $F_{K^\pm}/F_{\pi^\pm}$ where the electromagnetic radiative
corrections have been removed.

The experimental results of \cite{Rosselet} were already corrected for
radiative corrections. In the absence of better calculations and measurements
we therefore use our earlier procedure \cite{ABT2} again. We calculate this
process with the physical mass of the charged Kaon and pion.

\section{Other assumptions}
\label{assuptions}

The calculations in CHPT we have performed are all to \order{p^6}. We have
not used any improvement procedure to guess the size of the
higher order terms nor can we determine experimentally the \order{p^6}
tree terms.
At present the data are simply not constraining enough
to allow us to do so. The total number of
parameters is very large. Thus 
these have to be estimated
using a theoretical procedure as discussed in Sect. \ref{assumptions6}.
In addition, the available data and \order{p^6} calculations are not sufficient
to determine all other inputs, the assumptions we use here
are given in Sect. \ref{assumptions4}.

\subsection{At next-to-next-to-leading order, \order{p^6}}
\label{assumptions6}

When the full fits of the \order{p^4} LEC's became available \cite{GL}
an understanding of their general pattern and sizes within about 30\%
could be obtained using resonance saturation \cite{EGPR}. These results
were later refined in the framework of quark-models and large $N_c$
\cite{rest}.

We will use the former procedure, a few tests at \order{p^6} exist,
in two-point-functions some parameters could be determined
\cite{KG}
and in the pion form-factors two additional tests were possible \cite{BCT}.
All of these were successful within about 30\%. The precise
estimates we use are described in \cite{ABT1,ABT2}. We test this
assumption by varying the input in several ways\,:
\begin{enumerate}
\item Keeping only the two main contributions\,: vectors and $\eta^\prime$.
\item Varying the input parameters of the resonances by a factor of two.
\item Choosing different renormalization points where we do the estimates.
\end{enumerate}
The combination of the above 3 variations results in a rather large
variation of the \order{p^6} constants used. We thus expect that
this part does not affect our result within the quoted errors.

\subsection{At next-to-leading order, \order{p^4}}
\label{assumptions4}

Furthermore, using the data presented in \cite{Rosselet}
we need to fix four more inputs, we choose the quark mass ratio
$m_s/\hat{m}$, and three of the LEC's\,: $L_4^r$, $L_6^r$ and $L_9^r$. 
The new free parameter compared
to \cite{ABT2} is $m_u/m_d$ but now there are two Kaon masses instead
of one as fitting parameters so $m_u/m_d$ can be determined as a function of
the inputs.

For the quark mass ratio $m_s/\hat{m}$ we use as standard value $24$.
This is a reasonable average of the ratio of the sum-rule calculations of
$\hat{m}$ \cite{mhat} and $m_s$ \cite{mstrange}. We will also check
the changes of the fits with a rather broad range of inputs for
$m_s/\hat{m} \in [20,30]$.

The constant $L_9^r$ driving the pion charged radius
is saturated up to some good extent by
a Vector Dominance expression. The pole due to the $\rho$ exchange represents
the dominant low-energy singularity while the $\pi\pi$ and $K \bar{K}$ cuts
generate a tiny correction. 
This LEC is fixed at its standard value of $6.9\times 10^{-3}$. 
Even putting
it to zero hardly changes the results so this assumption does not bring
extra uncertainty.

The largest uncertainty w.r.t. the masses and decay constants is in the values
of $L_4^r$ and $L_6^r$. In the large $N_c$ limit they vanish and
varying the scale $\mu$ from 500~MeV to 1~GeV changes 
$L_4^r$ by about $\pm 0.5\times 10^{-3}$ 
and $L_6^r$ by about $\pm 0.3\times 10^{-3}$.
A range within zero of that size is thus an acceptable range.
There are some indications that the values are at the end of this region
and even somewhat outside for $L_6^r$ \cite{Bachir,DS}. 
We will make use of two experimental data analysis\,: in the first one
we will use 
\cite{Rosselet} (otherwise using 
\cite{pislak} barely will change the conclusions)
and will keep both of them, $L_4^r$ and $L_6^r$ at zero for our main
fit but discussing the variation of the results below. In second place we will
use the analysis of \cite{truol}, this will allow to remove one of the 
assumptions for the large $N_c$ behaviour of $L_4^r$ or $L_6^r$ and
compare the
result with the previous one.
Let us note here that there are possibly large corrections to the large $N_c$
limit in the scalar sector but they
are to a large extent due to the chiral loop
corrections and are thus included in the present results.
Examples are $\pi$-$\pi$ scattering \cite{pipi} and the pion scalar
form-factor \cite{BCT}. A large $N_c$ saturation of the constants in the latter
case provided excellent agreement with experiment. This question does
however deserve further study. But if we assume that the $SU(2)$
quantities should not be too sensitive to $m_s$ then we need to
assume that $L_4^r$ and $L_6^r$ are small. We also obtained \cite{ABT2} that
the $1/N_c$ suppressed combination $(2L_1^r-L_2^r)/L_3^r$ was small
providing one more test of the large $N_c$ assumption.
Let us comment briefly about the resonance saturation pattern
of the LEC.
The symmetry breaking part of the effective Lagrangian at \order{p^4}
is determined by $L_4^r,\,\ldots\,,L_{8}^r$. Something in common
to all of these constants is the fact that
in the low-energy regime none
is affected by the exchange of spin $1$ resonances but
of scalars and pseudoscalar particles, sector in which the experimental
data are still
fuzzy. Further more, for $L_4^r,\,\ldots\,,L_{8}^r$
the \emph{natural} scale ($M_S\simeq M_{\eta^{\prime}}\simeq 1
~\mbox{GeV}$)
differs from the one of $L_1^r, L_2^r$
and $L_3^r$, set in the last ones by Vector exchange ($M_V\simeq
M_\rho\simeq 0.8 ~\mbox{GeV}$). This obviously is reflected in the expansion
in powers of $m_s$, some quantities are controlled by the parameter
$(M_K/M_S)^2 \sim 1/9$ while others present a worse behaviour
$(M_K/M_\rho)^2 \sim 1/4$.

\section{$K_{\ell 4}$ decay parametrization and data}
\label{data}

The experiment E865 was not primarily designed to collect $K_{\ell 4}$
data but it recorded it as a background anyway~\cite{truol}.
These data were previously analyzed in a way similar to the older
experiment \cite{Rosselet} and presented in \cite{pislak}.
The second and latest analysis \cite{truol} uses a more suitable 
parametrization of the form-factors to fit the data at low-energy \cite{AB}
\ba
F&=&\{ f_s(s_\pi)+f_\ell s_\ell \} e^{i\delta^0_0(s_\pi)}
+ \mbox{p-wave}\,,\\ \nonumber
G&=&\{ g_p(s_\pi)+g_\ell s_\ell \} e^{i\delta^1_1(s_\pi)}
+ \mbox{d-wave}\,,
\ea
where
$K \to \pi^+(p_+)\, \pi^-(p_-)\, \ell(p_\ell)\, \nu(p_\nu)$ decay
variables are defined as
\eq
s_\ell= (p_\ell+p_\nu)^2\,, \quad
s_\pi = (p_++p_-)^2\,. 
\en
The analysis of \cite{truol} assumes only dependence in the lowest
partial wave decomposition, i.e., s-wave in $F$ and p-wave in $G$.
Furthermore any sensitivity on $s_\ell$ is discarded. As previous
work showed, \cite{ABT2}, those assumptions where valid in the
analysis of \cite{Rosselet}. But in view of the errors quoted in \cite{truol}
for the form-factor $F$ at threshold, $f_s(0)$,
this assumption is  borderline for the 
new data.

The form-factors in \cite{truol} where fitted to the following
expressions
\ba
F&=&(f_s + f_s^\prime s_\pi + f_s^{\prime\prime} s_\pi^2) 
e^{i\delta^0_0(s_\pi)}\,,
\\ \nonumber
G&=&(g_p + g_p^\prime s_\pi ) e^{i\delta^1_1(s_\pi)}\,,
\ena
where $f_s, f_s^\prime, f_s^{\prime\prime}, g_p$ and $g_p^\prime$
do not depend on any kinematical variable leading to the result
\ba
\label{fg}
f_s(0)&=&5.75\pm 0.02\pm 0.08\,,\quad f_s^\prime(0)=1.06\pm 0.10 \pm 0.40\,, 
\\ \nonumber
g_p(0)&=&4.66\pm 0.47 \pm 0.07\,,\quad g_p^\prime(0)=0.67 \pm 0.10 \pm 0.04\,. 
\ea
The quadratic slope is rather larger than we expect from our CHPT results
\cite{ABT2}. We therefore use the results of \cite{truol} with a linear
fit only
\ba
f_s(0)&=&5.770\pm0.097\,,\quad f_s^\prime(0)=0.47\pm0.15\,,
\\ \nonumber
g_p(0)&=&4.684\pm0.092
\,,\quad g_p^\prime(0)=0.54\pm0.20\,.
\ea
and regards these fits as preliminary till the data are finalized.

\section{The fitting procedure and fit results}
\label{fitting}

The fitting procedure is the same as the one used in \cite{ABT2} but
slightly adapted to the present case. 
We refer to \cite{ABT2} and the previous sections for a description of
the details about extra assumptions, besides CHPT,  
and how electromagnetic corrections were taken into account in all
the computed physical quantities.
Here we only sketch briefly the inputs and outputs of the calculations.

In addition to the lowest order Lagrangian
that contains: $B_0\,m_u\,,B_0\,m_d\,,B_0\,m_s$ and $F_0$ as free
constants (see below)
the \order{p^4}
Lagrangian is parametrized in terms of $10$ unknown LEC's not restricted by
symmetry\,: $L_1^r,\,\ldots\,,L_{10}^r$. All in all this amounts to
$14$ free parameters in total \footnote{Notice that
we have one more parameter than in \cite{ABT2} due to the isospin breaking.}.
The constant $L_{10}^r$ does not appear to
\order{p^6} in any of our quantities, thus remain $13$ free parameters
to be determined at \order{p^6}. Taking into account the restrictions
of Sect.~\ref{assuptions}
the number of parameters is reduced to $10$ or $9$ depending if we use
the results of
\cite{Rosselet} or \cite{truol} respectively. Hereafter we will quote the
quantities referring \cite{truol} between parenthesis.
The experimental inputs used are
four of the five pseudoscalar
masses, i.e. we discard the use of
$m_{\pi^\pm}^2$ because its difference with $m_{\pi^0}^2$ is purely of 
electromagnetic origin.
Two decay constants $F_{\pi^\pm}\,,F_{K^\pm}$. And \emph{three} (\emph{four})
inputs from $K_{\ell4}$ decays, Sect.~\ref{data}, 
two form-factors at threshold, 
$f_s(0)\,, g_p(0)$, and one (two) slope
$f^\prime_s(0)\,, (g^\prime_p(0))$.
In all there are nine (ten) inputs and
therefore this allows to determine nine (ten) parameters. 
Schematically
\begin{eqnarray}
\mbox{Inputs}&& \left\{ \begin{array}{l}
F_{\pi^\pm}, F_{K^\pm}   \vspace{.05in}  \\               
M_{\pi^0}^2\,,M_{K^\pm}^2\,,M_{K^0}^2\,,M_{\eta}^2  \vspace{.05in}  \\
f_s(0)\,,g_p(0)\,,f_s^\prime(0)\,,(g_p^\prime(0))
               \end{array}\right. 
\\
\mbox{Outputs}&& \left\{ \begin{array}{l}
F_0\vspace{.05in}  \\               
B_0\,\hat{m} \vspace{.05in} \\
m_u/m_d \vspace{.05in} \\
L_1\,,L_2\,,L_3\,,(L_4)\,,L_5\,,L_7\,,L_8                    
               \end{array}\right. 
\end{eqnarray}

Notice that using \cite{truol} we obtain 
between our outputs the order parameter $L_4^r$.
Unfortunately the $K_{\ell4}$ quantities are not very sensitive 
to $L_4^r$ \cite{Kl4oneloop}.

The errors used in the fit were the experimental errors on
$f_s\,,f_s^\prime$ and $g_p\,,(g_p^\prime)$, 
0.01 for $F_K/F_\pi$ and
we required the quark mass ratios determined from the lowest order
meson masses to be satisfied within 10\% accuracy.

The procedure for the latter was to take the physical meson mass,
calculate the \order{p^4} and \order{p^6} corrections to obtain the
lowest order masses of $\pi^\pm$, $\pi^0$, $K^\pm$, $K^0$ and $\eta$.
Since the lowest order masses actually were needed in the calculations,
they surface in the formulas via $\epsilon$ and $\Delta m^2$, we
iterated the procedure till it converged.

The change in the $L_i^r$ using \cite{Rosselet} are
compared to those reported in \cite{ABT2}
for all the fits was significantly smaller than the errors quoted there,
indicating that the neglect of isospin violation was a good
approximation\footnote{\label{footerror}Note that
the change in value is due to an error in the numerical program used
in \cite{ABT2} as described in the erratum.}.
The changes are always below $0.02\times 10^{-3}$ in magnitude.
We therefore still consider the main fit of \cite{ABT2} as the
standard values for the $L_i^r$. 

The remaining electromagnetic effects are estimated by also performing
a fit, this time
with the masses of all the propagating particles inside the loop
set to their physical mass rather than to the strong mass only. Again, the
changes in the results were rather small compared to the quoted errors.

\begin{table}[t]
\caption{\label{Tableresults:1}
Results for $L_i^r(\mu), F_0, B_0 \hat{m}$ and $m_u/m_d$ for the various
fits described in the main text using \cite{Rosselet}.
Errors are fitting errors as quoted by MINUIT.
All $L_i^r(\mu)$ values quoted have been brought to the scale
$\mu=0.77$~GeV. The first line with $m_u/m_d$ (full)
is with Eq. (\ref{BPdashen}).
while the second line, $m_u/m_d$ (simple) is with Dashen's theorem,
Eq. (\ref{dashen}).
The standard values are $m_s/\hat m = 24$, $\sqrt{s_\pi^\prime}= 0.336$~GeV,
 $s_\ell = 0$, $L_4^r=L_6^r=0$ and $L_9^r=6.9~10^{-3}$.}
\vspace{0.25cm}
\begin{small}
\begin{tabular*}{\textwidth}{*{11}{c@{\hspace{1.7mm}}}c}
\hline
  & Main Fit & \order{p^4}                     &fit 2  & fit 3 & fit 4 & fit 5 & fit 7 & fit 8 & fit 9\\
\hline
$10^3\,L^r_1$         &   0.53$\pm$0.25&   0.46&   0.53&   0.50&   0.50&   0.53&   0.43&   0.63&   0.66\\
$10^3\,L^r_2$         &   0.71$\pm$0.27&   1.49&   0.72&   0.66&   0.73&   0.80&   1.01&   0.72&   0.85\\
$10^3\,L^r_3$         &$-$2.72$\pm$1.12&$-$3.18&$-$2.73&$-$2.60&$-$2.76&$-$2.76&$-$2.90&$-$2.72&$-$3.32\\
$10^3\,L^r_5$         &   0.91$\pm$0.15& 1.46  &   0.87&   0.91&   0.90&   0.91&   1.48&   0.72&   0.86\\
$10^3\,L^r_7$         &$-$0.32$\pm$0.15&$-$0.49&$-$0.25&$-$0.32&$-$0.32&$-$0.32&$-$0.30&$-$0.30&$-$0.32\\
$10^3\,L^r_8$         &   0.62$\pm$0.20& 1.00  &   0.48&   0.62&   0.62&   0.63&   0.75&   0.56&   0.62\\
$F_0$ [MeV]           & 87.1           & 81.1  & 87.2  & 87.1  & 86.6  & 98.9  & 82.9  & 90.9  & 86.7  \\
$B_0\hat{m}$ [GeV$^2$]& 0.0136         & 0.0181&0.0141 &0.0136 &0.0136 &0.0138 &0.0158 & 0.0110& 0.0136\\
$\delta m_\pi$ [MeV]  & 0.32           & 0.27  & 0.16  & 0.32  & 0.32  & 0.23  &$-$0.07& 0.63  & 0.36  \\
$m_u/m_d$ (full)      &0.46$\pm$0.05   & 0.52  & 0.43  & 0.45  & 0.46  & 0.46  & 0.47  & 0.33  & 0.46  \\
$m_u/m_d$ (simple)    &0.52$\pm$0.05   & 0.58  & 0.50  & 0.53  & 0.53  & 0.53  & 0.54  & 0.43  & 0.53  \\
\hline
changed     & &${\cal O}(p^4)$&$m_s/\hat m$&$\sqrt{s_\ell}$&
$\sqrt{s_\pi^\prime}$&$L_4^r;L_6^r$ &$\mu$&$\mu$&$g(0)$\\
quantity     &          &       & 26 & 0.1             & 0.293
&$-0.3;-0.2$&   0.5&1.0&4.93\\
Unit         &          &       &    & GeV             & GeV   &
 $10^{-3}$&     GeV & GeV &\\
\hline
\end{tabular*}
\end{small}
\end{table}

The fit results using \cite{Rosselet}
are presented in Table~\ref{Tableresults:1}.
This can be compared directly with Table~2 in \cite{ABT2}\footnote{See footnote \ref{footerror}.}. 
With respect to this we have discarded
the comparison with previous works \cite{GL,old} (second column in Table~2 in
\cite{ABT2})
and fit 6, performed only with vectors and $\eta^\prime$
in the $C_i^r$ saturation, otherwise
we keep the same notation and only include $m_u/m_d$,
 $F_0$ and  $B_0 \hat{m}$ as output parameters.
We also performed the fit with varying the resonance input parameters
by a factor of two with similar results.

Let us summarize our findings in Table~\ref{Tableresults:1}\,: 
In the Main fit we use a strict
standard approach, considering the central value of the Zweig Rule violating
term, $L_4^r\simeq ~L_6^r=0$, $r=24$ and the scale is set to $M_\rho$. 
The second column compares each quantity with the corresponding
\order{p^4} results using the same input assumptions.
Fits 3 and 4 correspond to different choices in the kinematical points
for the $K_{\ell 4}$ variables.
In fit 5 we allow for a small violation of the Zweig Rule.
Fits 7 and 8 show different choices of the scale dependence $\mu$.
Our results are fully $\mu$-independent, but numerically
the change reflects our lack of knowledge about the scale where the
resonance saturation for the LEC's, $C_i^r$, is valid. 
Finally fit 9
presents a weighted
average of the p-wave form-factor $g_p(0)$ \cite{Rosselet,Makoff}.

There are no changes w.r.t. the numbers
in \cite{ABT2}\footnote{See footnote \ref{footerror}.}. Our results are therefore stable against
the isospin breaking effects. Furthermore the values of
$L_5^r$ and $L_8^r$ are compatible with the recent lattice
simulations \cite{latt}.

In order to show the changes introduced by the analysis of \cite{truol} we also
include Table~\ref{Tableresults:2}. Notice that the values 
quoted there are only \emph{indicative} and should be taken as preliminary till
the final E865 analysis is available. As explained in Section \ref{data} we use
the linear fit of \cite{truol} to perform the fit.

In Table~\ref{Tableresults:2} we have displayed the following results\,:
fit 10 is the direct equivalent of the main fit, and the next column
are the corresponding \order{p^4} values. 
Fit 11 is the equivalent of the main fit but leaving $L_4^r$ as a free 
parameter. As one can see there are no significant changes w.r.t. assuming
$L_4^r \sim 0$ and we only obtain a very weak limit on $L_4^r$.

This is not conclusive, i.e., one can not claim that
$L_6^r = 0 \Rightarrow L_4^r \sim 0$ is the only possible solution because
both constants can be highly correlated. For this reason we show
in fits 12 and 13 the shift due to different choice of $L_6^r$. 
As one can see from the last three columns of Table \ref{Tableresults:2}
the values of $L_6^r$ and $L_4^r$ follow almost a linear relation. Therefore
a big deviation from the Zweig Rule value of $L_6^r$ 
will probably signal also a deviation for $L_4^r$.
The possibility of a small value of $L_6^r$ and a large value of
$L_4^r$ fixed via the scalar pion radius is thus unlikely.

\begin{table}[t]
\caption{\label{Tableresults:2}
Results for $L_i^r(\mu), F_0, B_0 \hat{m}$ and $m_u/m_d$ for the various
fits described in the main text using the inputs of \cite{truol}.
$L_4^r$ and $L_6^r$ are given, $\equiv$ means it is an input value.
All other inputs are as in Table \ref{Tableresults:1}.}
\vspace{0.25cm}
\begin{center}
\begin{tabular}{c c c c c c}
\hline
       & fit10         & ${\cal O}(p^4)$& fit11          & fit12      & fit 13     \\
\hline
$10^3\,L^r_1$&   0.43$\pm$0.12  &   0.38&   0.43$\pm$0.12&   0.43     &   0.43     \\
$10^3\,L^r_2$&   0.73$\pm$0.12  &   1.59&   0.77$\pm$0.20&   0.85     &   0.70     \\
$10^3\,L^r_3$&$-$2.35$\pm$0.37  &$-$2.91&$-$2.36$\pm$0.40&$-$2.40     &$-$2.32     \\
$10^3\,L^r_4$&   $\equiv$0    &$\equiv$0&$-$0.18$\pm$0.85&$-$0.46     &   0.10     \\
$10^3\,L^r_5$&   0.97$\pm$0.11  & 1.46  &   1.08$\pm$0.61&   1.07     & 1.11       \\
$10^3\,L^r_6$&  $\equiv$0     &$\equiv$0&$\equiv$0       &$\equiv-0.2$&$\equiv 0.2$\\ 
$10^3\,L^r_7$&$-$0.31$\pm$0.14  &$-$0.49&$-$0.35$\pm$0.30&$-$0.35     &$-$0.36     \\
$10^3\,L^r_8$&   0.60$\pm$0.18  & 1.00  &   0.70$\pm$0.59&   0.70     & 0.70       \\
$F_0$ [MeV]  & 87.7             & 81.1  &   93.9         &  105.7     & 84.1       \\
$B_0 \hat{m}$ [GeV$^2$]& 0.0135 & 0.0181&   0.0104       & 0.0106     & 0.0104     \\
$\delta m_\pi$ [MeV]   & 0.28   & 0.27  & 0.83           & 0.53       & 1.12       \\
$m_u/m_d$ (full) &0.45$\pm$0.05 & 0.52  &   0.37$\pm$0.12& 0.38       & 0.38       \\
$m_u/m_d$ (simple)&0.52$\pm$0.05& 0.58  &   0.45$\pm$0.27& 0.45       & 0.47       \\
\hline
\end{tabular}
\end{center}
\end{table}

\section{Survey of Applications}
\label{oldres}
In this section we collect various results in the
light of the new values of the main fit in Table~\ref{Tableresults:1}.
We want to \emph{stress} that we do not study the convergence of CHPT here,
we only want to check that the total size of corrections is not very large
as compared to the lowest order.
The masses typically have small \order{p^4} corrections but
sizable \order{p^6} ones.
As we mentioned in \cite{ABT1,ABT2} there can be several reasons for this kind
of  
behaviour, we rephrase them here for sake of completeness\,:
there is a strong suppression between the LEC's and chiral logarithms at 
\order{p^4} and the assumption of resonance saturation using our naive
scalar picture at
\order{p^6} might also be incomplete.

\subsection{Meson masses}

The numerical results for the masses can be presented in various ways.
Using the definition Eq. (\ref{defmass}) we obtain
\ba
m_{\pi^\pm}^2/(m_{\pi^\pm}^2)_{\mbox{\tiny QCD}}
& = & 0.742 + 0.007 + 0.251\,,
\nonumber\\
m_{K^\pm}^2/(m_{K^\pm}^2)_{\mbox{\tiny QCD}}
& = & 0.693 + 0.027 + 0.280\,,
\nonumber\\
m_{\eta}^2/(m_{\eta}^2)_{\mbox{\tiny phys}}
& = & 0.741 - 0.029 + 0.289\,.
\ea
These results are the same within errors as those given in \cite{ABT2}.
Hereafter the first quoted number refers to the Born approximation, the
second to the next-to-leading correction and the third to the
next-to-next-to-leading contribution.
Notice, that as previous results concerning the masses, even if the \order{p^6}
contributions are bigger than the \order{p^4} all three appear
to be of the same size,
$\left(m_{\pi^\pm}^2/(m_{\pi^\pm}^2)_{\mbox{\tiny QCD}}\right)^{(6)}
\sim \left(m_{K^\pm}^2/(m_{K^\pm}^2)_{\mbox{\tiny QCD}}\right)^{(6)}
\sim \left(m_{\eta}^2/(m_{\eta}^2)_{\mbox{\tiny phys}}\right)^{(6)}$. 
Furthermore, almost all the quantities (besides the \order{p^6} LEC's)
contributing to the masses
have the same sign. Thus in order to make $\order{p^4} \ge \order{p^6}$ 
there must be some fine tuning of the \order{p^6} LEC's.   
Our simple \emph{not well controlled} estimate 
of the LEC's at \order{p^6} can thus be the reason for large
\order{p^6} corrections.

The isospin breaking quantities have as expansion
\ba
\left(m_{K^0}^2-m_{K^\pm}^2\right)_{\mbox{\tiny QCD}}
 &=& (5.09 - 0.34 +1.53)\times 10^{-3}    ~\mbox{GeV}^2\,,
\nonumber\\
\left(m_{\pi^\pm}^2-m_{\pi^0}^2\right)_{\mbox{\tiny QCD}}
 &=& (3.96 + 1.54 +3.17)\times 10^{-5}    ~\mbox{GeV}^2\,.
\ea

The Gell-Mann--Okubo relation in the presence of isospin breaking
is
\be
\Delta_{GMO}\equiv
m_\eta^2 + m_{\pi^0}^2 - \frac{2}{3}\left(m_{K^\pm}^2+m_{K^0}^2\right)
-\frac{2}{3}m_{\pi^\pm}^2 = 0\,.
\ee
The lowest order masses of course obey this identically and we use it
extensively throughout the calculation.
The deviation is given by
\be
\Delta_{GMO}
= (0-0.0172-0.0033)~\mbox{GeV}^2 =  -0.0205~\mbox{GeV}^2\,.
\ee

\subsection{Decay constants}

The decay constant have as expansion
\ba
F_{\pi^\pm}/F_0 &=& 1 + 0.135 -0.075\,,
\nonumber\\
F_{K^\pm}/F_{\pi^\pm} &=& 1 + 0.162 +0.058\,,
\nonumber\\
F_{\eta8}/F_{\pi^\pm} &=& 1.000 + 0.242 + 0.066 = 1.308\,,
\ea
and for the isospin violating quantities
\ba
\frac{F_{\pi^03}-F_{\pi^\pm}}{F_{\pi^\pm}}
& = & (1.0 +0.3 + 0.8)\times 10^{-4}\,, 
\nonumber\\
\epsilon_\pi = \frac{F_{\pi^08}}{F_{\pi^03}}&=&
0.0141+0.0039-0.0021 =  0.0159\,,
\nonumber\\
\epsilon_\eta = -\frac{F_{\eta3}}{F_{\eta8}}&=&
0.0141-0.0010-0.0026 =  0.0105\,.
\ea

\subsection{Vacuum expectation values}

The effective Lagrangian at lowest order contains two order parameters
$F_0$ and $B_0$ both of them sensitive to the infrared
end of the Dirac spectrum \cite{smilga} but
their r\^ole in the spontaneous chiral symmetry breaking pattern
is quite different.
The only necessary and sufficient condition of SB$\chi$S is a nonzero
value of the correlator function
\be
\lim_{m\to 0} i \int d^4x 
\langle\,\Omega\,\vert\,:L^i_\mu(x) R^j_\nu(0):\,\vert
\,\Omega\rangle = -\frac{1}{4}\eta_{\mu\nu} \delta^{ij} F_0^2
+\order{m_q}\,, 
\ee
where $L_\mu=\frac{1}{2}(V_\mu-A_\mu)$ and $R_\mu=\frac{1}{2}(V_\mu+A_\mu)$ 
are Noether currents. There is no such constraint on $B_0$
whose size is determined by the vacuum expectation value
\be
\label{vac}
\langle\,0\,\vert\,\bar{q}^i q^j\,\vert\,0\,\rangle =-F_0^2 B_0 \delta^{ij}
\order{m_q}\,,
\ee
and in principle has no lower bound other than vacuum stability,
$B_0 \geq 0$. Even though there is no proof available is rather likely that
$QCD$ is not realized in the phase $B_0=0$ \cite{sterntalk}.

The chiral
expansion of the vacuum expectation value (\ref{vac}) can be ordered as
\be
\langle\,0\,\vert\,\bar{q}^i q^j\,\vert\,0\,\rangle = - F_0^2 B_0
\{1+\langle\,0\,\vert\,\bar{q}^i q^j\,\vert\,0\,\rangle^{(4)}+
\langle\,0\,\vert\,\bar{q}^i q^j\,\vert\,0\,\rangle^{(6)}
+\ldots\}\,, 
\ee
where the superscripts refer to the chiral order.
Although  the result is scale
independent order by order in the CHPT expansion
it depends on the
$QCD$ renormalization scale $\mu_{QCD}$.
It follows from the fact that the quark scalar current needs to be defined
precisely in QCD. Already at \order{p^2} $B_0$ depends on $\mu_{QCD}$.
At \order{p^4} there appears an additional ambiguity
via the \emph{high} energy constant $H_2$ which 
 is forced by the ambiguity in the subtraction
of the scalar two-point function
\be
\langle 0 \vert \bar{q}q \vert 0 \rangle =
\langle 0 \vert \bar{q}q \vert 0 \rangle \Big\vert_{\mbox{\small chiral~limit}}
 - i \int d^4x \langle 0 \vert :(\bar{q}{\cal M} q)(x)(\bar{q}q)(0):
 \vert 0\rangle+{\cal O}({\cal M}^2)\,,
\ee
at zero distance.
At \order{p^4} one can eliminate the constant $H_2$ and thus obtain a 
sum rule relating the ratios of the isospin asymmetry
$\langle 0 \vert \bar{d}d \vert 0 \rangle /
\langle 0 \vert \bar{u}u \vert 0 \rangle$
to the $SU(3)$ asymmetry
$\langle 0 \vert \bar{s}s \vert 0 \rangle /
\langle 0 \vert \bar{u}u \vert 0 \rangle$. 
In order to quote numbers we fix $H_2$ assuming scalar dominance
\be
H_2^r = 2 L_8^r\,.
\ee
With this approach we obtain the following expansion
\ba
\langle 0 \vert \bar{u}u \vert 0 \rangle &=&
-B_0 F_0^2 [1+0.271+0.106 ]\,, \nonumber\\
\langle 0 \vert \bar{d}d \vert 0 \rangle &=&
-B_0 F_0^2 [1+0.284+0.110    ]\,, \nonumber\\
\langle 0 \vert \bar{s}s \vert 0 \rangle &=&
-B_0 F_0^2 [1+0.964+0.289 ]\,, 
\ea
where the quoted numbers correspond to the lowest order, \order{p^4}
and \order{p^6} respectively. The dependence on $H_2^r$ can be judged
from\footnote{The \order{p^6} dependence on $H_2^r$ is from rewriting
$F_0^{-2}$ in $F_{\pi^\pm}^{-2}$ in $\langle 0 \vert\bar{q}q\vert 0 \rangle$.}
\ba
H_2^r &=& 0\,,
\nonumber\\
\langle 0 \vert \bar{u}u \vert 0 \rangle &=&
-B_0 F_0^2 [1+0.264+0.104 ]\,, \nonumber\\
\langle 0 \vert \bar{d}d \vert 0 \rangle &=&
-B_0 F_0^2 [1+0.271+0.107  ]\,, \nonumber\\
\langle 0 \vert \bar{s}s \vert 0 \rangle &=&
-B_0 F_0^2 [1+0.693+0.222 ]\,, 
\ea
The vacuum asymmetries are defined as
\be
\frac{\langle\,0\,\vert\,\bar{d} d\,\vert\,0\,\rangle}
{\langle\,0\,\vert\,\bar{u} u\,\vert\,0\,\rangle} \equiv 1 -\epsilon_d\,,\quad
\frac{\langle\,0\,\vert\,\bar{s} s\,\vert\,0\,\rangle}
{\langle\,0\,\vert\,\bar{u} u\,\vert\,0\,\rangle} \equiv 1 -\epsilon_s\,.
\ee
Using Partially Conserved Axial Currents (PCAC) these ratios are
fixed to be $1$, then $\epsilon_d$ and $\epsilon_s$ measure
the $SU(2)$ and $SU(3)$ breaking of the non-perturbative vacuum respectively. 
We get for them
\ba
\label{ep}
\epsilon_d &=& -0.0129-0.0011\,,           \nonumber\\
\epsilon_s &=& -0.693+0.004\,,
\ea
where the first number corresponds to \order{p^4} and the second
to \order{p^6}. As becomes clear the breaking inside the $SU(2)$ doublet
driven by $\epsilon_d$ is very small while the breaking due to the 
presence of $m_s$, $\epsilon_s$, is 
rather huge.
There is a plethora of phenomenological calculations 
determining Eq.~(\ref{ep}) from sum-rules.
Their results are very spread and no clear conclusion
can be reached. Otherwise the results in Eq.~(\ref{ep}) are in the ball park
of those quoted in \cite{shi} for $\epsilon_d$ and \cite{un} for $\epsilon_s$
while the remaining results \cite{reste} are quite off of Eq.~(\ref{ep}). 

\section{Quark Mass Ratios}
\label{ratios}

A review of the early quark mass determinations can be found in
\cite{GLPhysRep}. A more recent list of references is the quark mass
section in the Review of Particle Properties \cite{PDG2000}.

While sum-rule or lattice techniques allow to extract the value
of quark masses via some hadronic observable, the use of an effective
Lagrangian, as CHPT, \emph{only} allows to determine the relative size
of the quark masses, $m_u, m_d$ and $m_s$.
This approach assumes that the quark masses
can be treated as a perturbation on the massless QCD Hamiltonian, 
${\cal H}_0$
\be
{\cal H}_{QCD} = {\cal H}_0 -
 \int\,d^3 x\, [m_u \bar{u} u+m_d \bar{d} d+m_s \bar{s} s ]\,.
\ee
Quark masses depend on the renormalization scale, but if we use
a mass-independent renormalization scheme like $\overline{MS}$
they are renormalized multiplicatively and their ratios
are renormalization-scale-independent.
Due to the locality of the symmetries in the chiral Lagrangian
the quark masses in CHPT can be identified with those in QCD via the
Ward identities, thus by construction they are 
the same as those in QCD.

The values of the relevant LEC's can depend on the precise definition
of the scalar currents in QCD. In practice, since we have to determine
the LEC's from the same data as we use to obtain the quark masses
some ambiguities might occur \cite{KM}. These ambiguities
are discussed in \cite{Heiri} where we refer for a more detailed
discussion. In the present context they
are fixed by our large $N_c$ assumptions on the values of
$L_4^r$ and $L_6^r$.

To lowest order in the chiral expansion and neglecting electromagnetic 
corrections we have
\ba
\label{massLO}
m_{\pi^0}^2-\delta &=& m_{\pi^\pm}^2 = B_0 (m_u+m_d) \nonumber\,, \\
m_{K^\pm}^2&=&B_0 (m_u+m_s) \nonumber\,, \\
m_{K^0}^2&=&B_0 (m_d+m_s) \nonumber\,, \\
m_\eta^2 &=& \frac{1}{3}B_0 (m_u+m_d+4 m_s)+\delta\,,
\ea
where $\delta$ is the effect of $\pi^0$-$\eta$ mixing. The constant
of proportionality is determined by the quark condensate
$B_0 = \vert\,\langle\,0\,\vert\bar{u}u\vert\,0\,\rangle\,\vert/F_\pi^2
+\order{m_q}$.
We can use Eq. (\ref{massLO}) to get
\ba
\label{LOratios}
\frac{m_s}{\hat m} &=&
\frac{m_{K^\pm}^2+m_{K^0}^2-m_{\pi^\pm}^2}{m_{\pi^\pm}^2} \approx  
\mbox{ 24.2 (real); 25.8 [Eq. (\ref{dashen})]; 25.7 [Eq. (\ref{BPdashen})]}\,,
\nonumber\\
\frac{m_u}{m_d}    &=&
\frac{m_{K^\pm}^2-m_{K^0}^2+m_{\pi^\pm}^2}
{m_{K^0}^2-m_{K^\pm}^2+m_{\pi^\pm}^2}\approx
\mbox{ 0.66 (real); 0.56 [Eq. (\ref{dashen})]; 0.49 [Eq. (\ref{BPdashen})]}\,.
\ea
The three quoted numbers correspond to using the physical masses,
electromagnetic corrections using Dashen's theorem, Eq. (\ref{dashen}),
 and $e^2m_s$ corrections included as well, Eq.~(\ref{BPdashen}).

To \order{p^4} we do not have quite enough information anymore
to determine both quark mass ratios and all the relevant LEC's.
To first order in isospin breaking it was noted in
\cite{GL,Heiri} that the particular combination
\be
\label{defQ}
Q^2 \equiv
 \frac{m_s^2-\hat{m}^2}{m_d^2-m_u^2}
\equiv \frac{1}{4}\left(1+\frac{m_s^2}{\hat m^2}\right)
\frac{1+(m_u/m_d)}{1-(m_u/m_d)}
=
\frac{m_K^2}{m_\pi^2} \frac{m_K^2-m_\pi^2}{m_{K^0}^2-m_{K^\pm}^2}
\ee
could be related to the meson masses without
corrections at this chiral order, thus providing a tight 
constraint on the quark mass ratio. As was pointed out in \cite{KM}
if the terms $\hat{m}^2/m_s^2$ are discarded Eq.~(\ref{defQ}) leads to an 
ellipsis where the variables are the ratios $m_u/m_d$ and $m_s/m_d$. The major 
semi-axis is given by $Q$ while the minor one is equal to $1$.
Plugging  the masses in Eq.~(\ref{defQ})
gives
\be
\label{Qp4}
Q = 22.0\, (21.3)
 ~\mbox{[Eq. (\ref{BPdashen})]}
;\quad
24.1\, (23.4)~ \mbox{[Eq. (\ref{dashen})]}\,.
\ee
The numbers are by using the strong meson masses in the last
part of Eq.~(\ref{defQ}).
The presence of higher orders corrections 
can be seen from the number in brackets, these are calculated 
using the quark mass ratios from
the fit with $m_s/\hat m=24$ at \order{p^4} of Table~\ref{Tableresults:1}
and the definition of $Q$ in terms of the quark masses.
Again the two sets of numbers correspond to using the
electromagnetic corrections 
estimated with $e^2m_s$ corrections included, Eq.~(\ref{BPdashen}),
or \order{e^2} only
using Dashen's theorem, Eq. (\ref{dashen}).

The quantity $Q$ can be determined 
to the same order from $\eta\to3\pi$
decays giving
\be
\label{Qeta}
Q = 22.7\pm0.8\,.
\ee
The decay
$\eta\to3\pi$ is known in CHPT to \order{p^4} \cite{GLeta} and higher order
corrections were estimated using dispersion relations \cite{etadispersive}.
In view of the discrepancy between dispersive estimates \cite{old}
and the full calculation to \order{p^6} \cite{ABT2} in the $K_{\ell4}$ process
the error in Eq. (\ref{Qeta}) is probably somewhat underestimated. 
The value is
also obtained using the average value of $\Gamma(\eta\to\gamma\gamma)$ of
\cite{PDG2000}.

\begin{figure}
\includegraphics[height=0.49\textwidth,angle=-90]{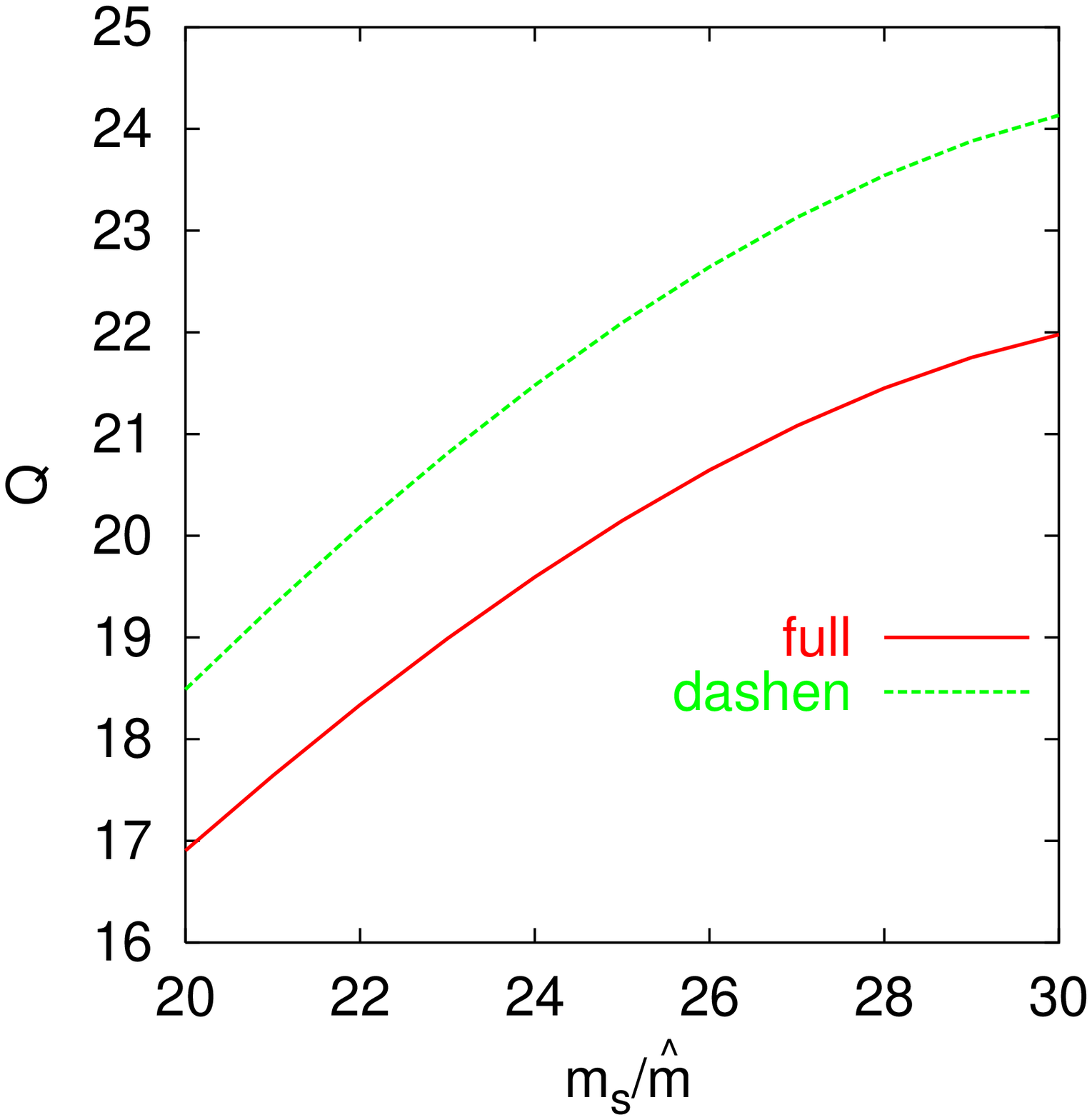}
\includegraphics[height=0.49\textwidth,angle=-90]{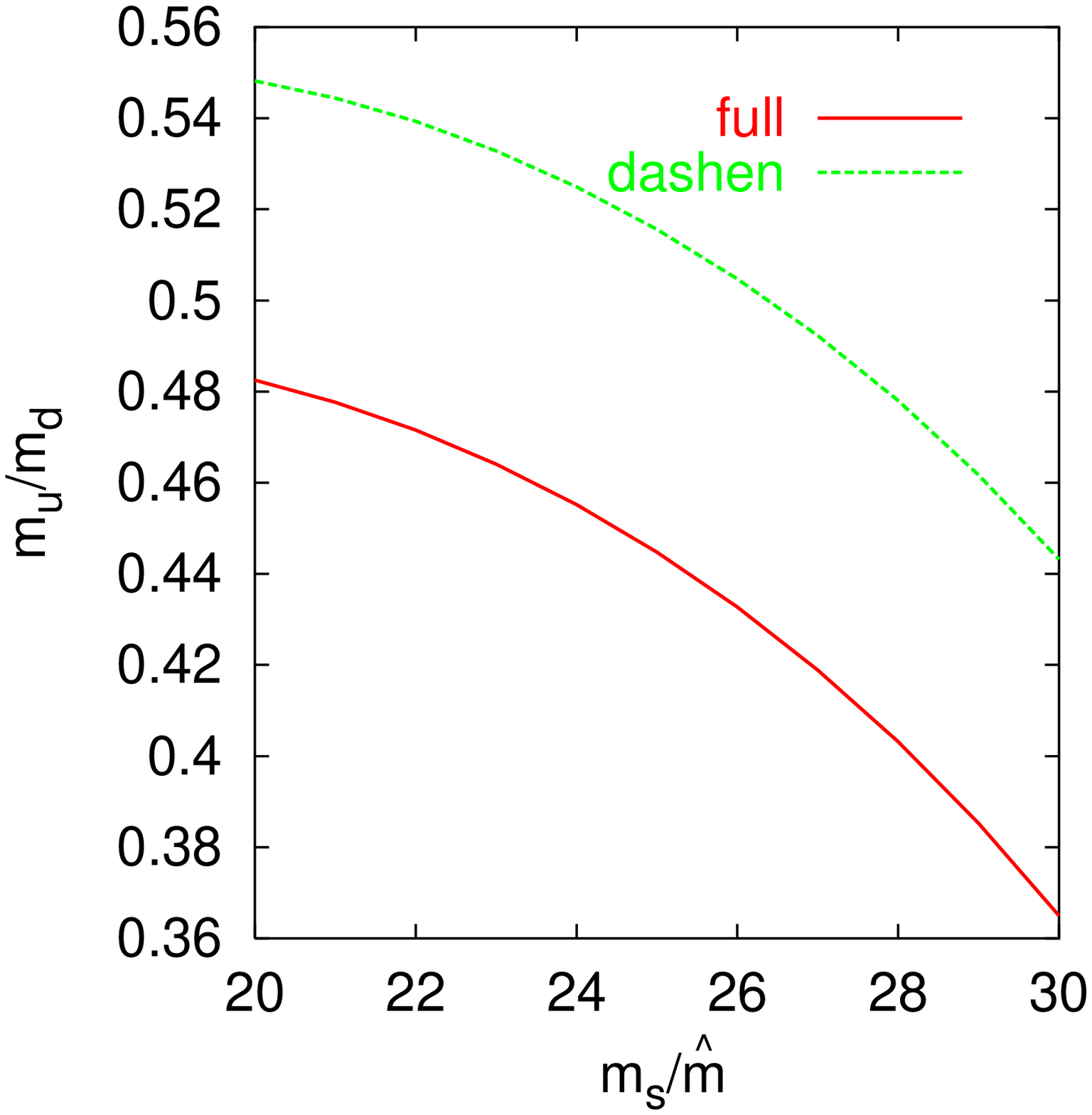}
\caption{\label{figmsmhat} (a) $Q$  (b) $m_u/m_d$
as a function of $m_s/\hat m$ but other input fixed as in the main fit.}
\end{figure}

Let us now turn to the \order{p^6} results. Our main fit with
$m_s/\hat m=24 $ gives
\be
\label{Q}
Q = 19.6 ~\mbox{[Eq. (\ref{BPdashen})]}\,;
\quad 21.5 ~ \mbox{[Eq. (\ref{dashen})]}\,.
\ee
substantially lower than the values quoted in Eq. (\ref{Qp4}).
On the other hand, $Q$ is rather sensitive to the input value for
$m_s/\hat m$. In Fig.~\ref{figmsmhat}(a) we have shown how $Q$
depends on the input values chosen for $m_s/\hat m$. Otherwise the inputs
are as in our main fit. Fig.~\ref{figmsmhat}(b) shows similarly
the dependence of $m_u/m_d$ on $m_s/\hat m$.

The position of the previous mentioned ellipsis 
can be described as the ratio of $SU(3)$ breaking effects
versus $SU(2)$ isospin breaking using
\be
R \equiv \frac{m_s-\hat m}{m_d-m_u}\,.
\ee
The standard value \cite{GLPhysRep} used $\rho$-$\omega$ mixing
and combinations of the baryon masses to obtain
$R=43.5\pm2.2$. Both of these inputs are now known to have rather
large corrections of \order{m_q}. $\rho$-$\omega$ mixing is treated
in \cite{BGT} and results for the baryon masses with earlier references can
be found in \cite{BM}. We obtain for this quantity
\eq
R\approx 31~\mbox{[Eq. (\ref{BPdashen})]}\,; 
\quad37 ~ \mbox{[Eq. (\ref{dashen})]}\,. 
\en

One of the main results 
from Table \ref{Tableresults:1} is 
\be
\label{mumdresult}
\frac{m_u}{m_d} = 0.46\pm0.09\,.
\ee
In the previous estimate the error is increased w.r.t. 
the one of the fit in order to include
two other sources of errors, about 0.02 from the error on the electromagnetic
correction due to $m_{K^\pm}^2-m_{K^0}^2$ and another 0.02 from the rather
large change that happened with the scale variation at $\mu=1$~GeV.
The other new result we obtain is the strong mass difference
\be
\delta m_\pi = \left(m_{\pi^\pm}-m_{\pi^0}\right)_{\mbox{\tiny QCD}}
= 0.32\pm0.20~\mbox{MeV}\,.
\ee
The change with the result quoted in \cite{GL} has two sources,
about 0.04~MeV from the inclusion of the mass-corrections in the
Kaon electromagnetic mass difference and the remainder comes
from the \order{p^6}
effects.

\begin{figure}
\label{mumd}
\begin{center}
\includegraphics[width=0.48\textwidth,angle=-90]{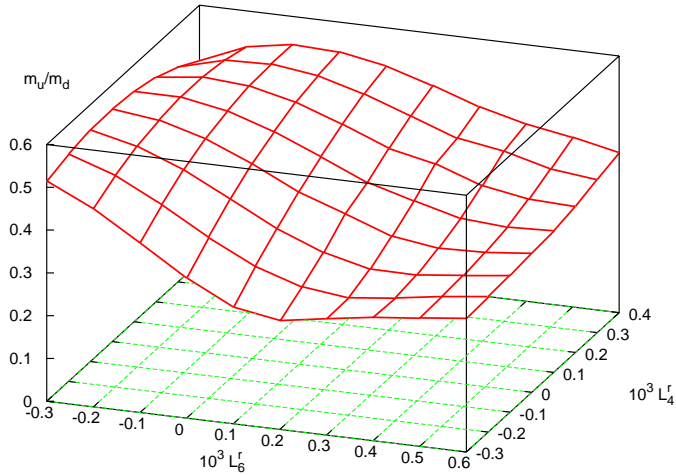}
\end{center}
\caption{\label{figmumd} The ratio $m_u/m_d$ versus
the inputs for $L_4^r, L_6^r$ the rest of parameters correspond to those
of our main fit. 
}
\end{figure}

A possibly more serious variation with input is due to
the assumptions on $L_4^r$ and $L_6^r$. Our tests of large $N_c$ as described
in Sect.~\ref{assumptions4} are mainly in quantities dominated by vector
exchange. We have therefore checked what happens when we vary
the input value of $L_4^r$ and $L_6^r$ over a fairly wide range.
The resulting values for $m_u/m_d$ are plotted in Fig.~\ref{figmumd}.
It should be kept in mind that not all of the points shown have good fits
to all inputs and most of the points actually fall in the
range of Eq.~(\ref{mumdresult}).
It is a rather clear conclusion that the value $m_u/m_d = 0$
is far away from the numbers we have obtained. Unfortunately this means that
the solution to the strong CP problem needs to be found elsewhere.

\section{Summary}
\label{sum}

In this paper we have explained how isospin breaking effects can be
incorporated at next-to-next-to-leading order in CHPT.
The expectation that isospin breaking would not significantly
alter our previous results has been confirmed and the results
presented here are
 a first step in understanding how isospin breaking
effects can be added consistently.

We have reanalyzed the main fit at \order{p^6} of the CHPT LEC's in the
presence of isospin breaking. The determination of the $L_i^r$ is
only marginally changed compared to the isospin symmetric analysis.
As by product we obtain control on the quantity
$m_u/m_d$ at next-to-next-to-leading order in terms of meson masses.
This constitutes one of our main new results.

Furthermore using the preliminary new experimental results
on $K_{\ell4}$ we can obtain one of the
Zweig Rule suppressed LEC, namely $L_4^r$.
It follows approximately a linear relation
together with $L_6^r$. 

We also revised the value of the isospin
breaking quantities $Q$ and $R$
discussing the electromagnetic effects. In
view of the low value of Eq.~(\ref{Q}) a reanalysis
of the $\eta \to 3 \pi$ decay at the same order in the quark mass expansion
as was done here seems necessary.

\vspace{1cm}

{\bf Acknowledgements}\\
G.A and P.T. were supported by TMR, EC--Contract No. ERBFMRX--CT980169 
(EURODAPHNE).\\


\begin{thebibliography}{99} 

\bibitem{GL}
J.~Gasser and H.~Leutwyler,
Nucl.\ Phys.\  {\bf B250} (1985) 465.

\bibitem{ABT1}     
G.~Amoros, J.~Bijnens and P.~Talavera,
Nucl.\ Phys.\  {\bf B568} (2000) 319
[hep-ph/9907264].

\bibitem{ABT2}
G.~Amoros, J.~Bijnens and P.~Talavera,
Phys.\ Lett.\  {\bf B480} (2000) 71
[hep-ph/9912398];
Nucl.\ Phys.\  {\bf B585} (2000) 293
[hep-ph/0003258].

\bibitem{roos}
H.~Leutwyler and M.~Roos,
Z.\ Phys.\  {\bf C25} (1984) 91.

\bibitem{truol}
P.~Truol {\it et al.}  [E865 Collaboration],
hep-ex/0012012.

\bibitem{CHPTlectures}
G.~Ecker,
``Strong interactions of light flavours,''
hep-ph/0011026;
A.~Pich,
``Effective field theory,''
hep-ph/9806303.

\bibitem{helmutmixing} 
G.~Ecker, G.~Muller, H.~Neufeld and A.~Pich,
Phys.\ Lett.\  {\bf B477} (2000) 88
[hep-ph/9912264].

\bibitem{pipi}
J.~Bijnens, G.~Colangelo, G.~Ecker, J.~Gasser and M.~E.~Sainio,
Nucl.\ Phys.\  {\bf B508} (1997) 263
[hep-ph/9707291].

\bibitem{Urech}
R.~Urech,
Nucl.\ Phys.\  {\bf B433} (1995) 234
[hep-ph/9405341].

\bibitem{KNRT}
M.~Knecht, H.~Neufeld, H.~Rupertsberger and P.~Talavera,
Eur.\ Phys.\ J.\  {\bf C12} (2000) 469
[hep-ph/9909284] and work in progress.

\bibitem{Dashen}
R.~Dashen,
Phys.\ Rev.\ {\bf 183} (1969) 1245.

\bibitem{NR}
H.~Neufeld and H.~Rupertsberger,
Z.\ Phys.\  {\bf C68} (1995) 91;
Z.\ Phys.\  {\bf C71} (1996) 131
[hep-ph/9506448].

\bibitem{hans}
J.~Bijnens,
Phys.\ Lett.\  {\bf B306} (1993) 343
[hep-ph/9302217].

\bibitem{dono}
J.~F.~Donoghue, B.~R.~Holstein and D.~Wyler,
Phys.\ Rev.\  {\bf D47} (1993) 2089.

\bibitem{lat}
A.~Duncan, E.~Eichten and H.~Thacker,
Phys.\ Rev.\ Lett.\  {\bf 76} (1996) 3894
[hep-lat/9602005];
Nucl.\ Phys.\ Proc.\ Suppl.\  {\bf 53} (1997) 295
[hep-lat/9609015].

\bibitem{BP}
J.~Bijnens and J.~Prades,
Nucl.\ Phys.\  {\bf B490} (1997) 239
[hep-ph/9610360].

\bibitem{BaurUrech}
R.~Baur and R.~Urech,
Phys.\ Rev.\  {\bf D53} (1996) 6552
[hep-ph/9508393].

\bibitem{Holstein}
B.~R.~Holstein,
Phys.\ Lett.\  {\bf B244} (1990) 83.

\bibitem{Finkemeier}
M.~Finkemeier,
Phys.\ Lett.\  {\bf B387} (1996) 391
[hep-ph/9505434].

\bibitem{Rosselet}
L.~Rosselet {\it et al.},
Phys.\ Rev.\  {\bf D15} (1977) 574.

\bibitem{EGPR}
G.~Ecker, J.~Gasser, A.~Pich and E.~de Rafael,
Nucl.\ Phys.\  {\bf B321} (1989) 311;
G.~Ecker et. al.,
Phys.\ Lett.\  {\bf B223} (1989) 425.

\bibitem{rest}
D.~Espriu, E.~de Rafael and J.~Taron,
Nucl.\ Phys.\  {\bf B345} (1990) 22;
J.~Bijnens, C.~Bruno and E.~de Rafael,
Nucl.\ Phys.\  {\bf B390} (1993) 501
[hep-ph/9206236];
J.~Bijnens,
Phys.\ Rept.\  {\bf 265} (1996) 369
[hep-ph/9502335];
S.~Peris, M.~Perrottet and E.~de Rafael,
JHEP {\bf 9805} (1998) 011
[hep-ph/9805442].

\bibitem{KG}
E.~Golowich and J.~Kambor,
Nucl.\ Phys.\  {\bf B447} (1995) 373
[hep-ph/9501318];
Phys.\ Rev.\  {\bf D58} (1998) 036004
[hep-ph/9710214];
S.~Durr and J.~Kambor,
Phys.\ Rev.\ D {\bf 61} (2000) 114025
[hep-ph/9907539].

\bibitem{BCT}
J.~Bijnens, G.~Colangelo and P.~Talavera,
JHEP {\bf 9805} (1998) 014
[hep-ph/9805389].

\bibitem{mhat}
C.~A.~Dominguez and E.~de Rafael,
Annals Phys.\  {\bf 174} (1987) 372;
J.~Bijnens, J.~Prades and E.~de Rafael,
Phys.\ Lett.\  {\bf B348} (1995) 226
[hep-ph/9411285];
S.~Narison,
Nucl.\ Phys.\ Proc.\ Suppl.\  {\bf 86} (2000) 242
[hep-ph/9911454].

\bibitem{mstrange}
K.~G.~Chetyrkin, C.~A.~Dominguez, D.~Pirjol and K.~Schilcher,
Phys.\ Rev.\  {\bf D51} (1995) 5090
[hep-ph/9409371];
M.~Jamin and M.~Munz,
Z.\ Phys.\  {\bf C66} (1995) 633
[hep-ph/9409335];
K.~G.~Chetyrkin, D.~Pirjol and K.~Schilcher,
Phys.\ Lett.\  {\bf B404} (1997) 337
[hep-ph/9612394];
P.~Colangelo, F.~De Fazio, G.~Nardulli and N.~Paver,
Phys.\ Lett.\  {\bf B408} (1997) 340
[hep-ph/9704249];
K.~Maltman,
Phys.\ Lett.\  {\bf B462} (1999) 195
[hep-ph/9904370];
A.~Pich and J.~Prades,
JHEP{\bf 9910} (1999) 004
[hep-ph/9909244];
J.~Kambor and K.~Maltman,
Phys.\ Rev.\ D {\bf 62} (2000) 093023
[hep-ph/0005156].

\bibitem{Bachir}
B.~Moussallam,
Eur.\ Phys.\ J.\  {\bf C14} (2000) 111
[hep-ph/9909292].

\bibitem{DS}
S.~Descotes and J.~Stern,
Phys.\ Lett.\  {\bf B488} (2000) 274
[hep-ph/0007082];
S.~Descotes,
hep-ph/0012221.

\bibitem{pislak}
S.~Pislak, 
in {\it 
talk given at Laboratori Nazionali di Frascati , 
Rome, Italy, June 22, 2000}, unpublished.

\bibitem{AB}
G.~Amoros and J.~Bijnens,
J.\ Phys.\ {\bf G25} (1999) 1607
[hep-ph/9902463].

\bibitem{Kl4oneloop}
J.~Bijnens,
Nucl.\ Phys.\ {\bf B337} (1990) 635;
C.~Riggenbach {\it et al.},
Phys.\ Rev.\ D {\bf 43} (1991) 127.

\bibitem{old}
J.~Bijnens, G.~Colangelo and J.~Gasser,
Nucl.\ Phys.\  {\bf B427} (1994) 427
[hep-ph/9403390].

\bibitem{Makoff}
G.~Makoff {\it et al.},
Phys.\ Rev.\ Lett.\  {\bf 70} (1993) 1591;
Erratum-ibid. {\bf 75} (1995) 2069.

\bibitem{latt}
J.~Heitger, R.~Sommer and H.~Wittig  [ALPHA Collaboration],
Nucl.\ Phys.\  {\bf B588} (2000) 377
[hep-lat/0006026].

\bibitem{smilga}
H.~Leutwyler and A.~Smilga,
Phys.\ Rev.\  {\bf D46} (1992) 5607.

\bibitem{sterntalk} 
J.~Stern,
[hep-ph/9801282] and  
in {\it 
talk given at Workshop on Chiral Dynamics: Theory and Experiment (ChPT 97), 
Mainz, Germany,
1-5 Sep 1997. 
}
[hep-ph/9712438].

\bibitem{shi}
M.~A.~Shifman, A.~I.~Vainshtein and V.~I.~Zakharov,
Nucl.\ Phys.\ {\bf B147} (1979) 519.

\bibitem{un}
S.~Narison,
Phys.\ Lett.\ {\bf B104} (1981) 485;
E.~Bagan, A.~Bramon, S.~Narison and N.~Paver,
Phys.\ Lett.\ {\bf B135} (1984) 463.

\bibitem{reste}
P.~Pascual and R.~Tarrach, 
Phys.\ Lett.\ {\bf B116} (1982) 443;
L.~J.~Reinders, H.~R.~Rubinstein and S.~Yazaki,
Phys.\ Lett.\ {\bf B120} (1983) 209. Erratum-ibid. {\bf B122} (1983) 487.

\bibitem{GLPhysRep}
J.~Gasser and H.~Leutwyler,
Phys.\ Rept.\  {\bf 87} (1982) 77.

\bibitem{PDG2000}
D.~E.~Groom {\it et al.},
Eur.\ Phys.\ J.\  {\bf C15} (2000) 1.

\bibitem{KM}
D.~B.~Kaplan and A.~V.~Manohar,
Phys.\ Rev.\ Lett.\  {\bf 56} (1986) 2004.

\bibitem{Heiri}
H.~Leutwyler,
``Non-lattice determinations of the light quark masses,''
hep-ph/0011049;
Phys.\ Lett.\  {\bf B378} (1996) 313
[hep-ph/9602366].

\bibitem{GLeta}
J.~Gasser and H.~Leutwyler,
Nucl.\ Phys.\  {\bf B250} (1985) 539.

\bibitem{etadispersive}
J.~Kambor, C.~Wiesendanger and D.~Wyler,
Nucl.\ Phys.\  {\bf B465} (1996) 215
[hep-ph/9509374];
A.~V.~Anisovich and H.~Leutwyler,
Phys.\ Lett.\  {\bf B375} (1996) 335
[hep-ph/9601237].

\bibitem{BGT}
J.~Bijnens, P.~Gosdzinsky and P.~Talavera,
Nucl.\ Phys.\  {\bf B501} (1997) 495
[hep-ph/9704212].

\bibitem{BM}
B.~Borasoy and U.~Meissner,
Annals Phys.\  {\bf 254} (1997) 192
[hep-ph/9607432].

\end{thebibliography}
\end{document}